\documentclass[twoside,twocolumn]{article}

\usepackage[sc]{mathpazo} 
\usepackage[T1]{fontenc} 
\usepackage{lmodern} 
\usepackage{microtype} 

\usepackage[english]{babel} 

\usepackage{graphicx}
\usepackage{color}
\usepackage[top=2cm,bottom=2cm,left=1.5cm,right=1.5cm,columnsep=20pt]{geometry} 
    {\large\scshape Computational Details%
    \par\medskip\normalfont\normalsize}%
    {}%
\newenvironment{acknowledgement}
    {\large\scshape Acknowledgement%
    \par\medskip\normalfont\normalsize}%
    {}%
    {\large\scshape Supporting Information%
    \par\medskip\normalfont\normalsize}%
    {}%
\usepackage{multirow} 
\usepackage{slashbox} 

\usepackage[normal, small,labelfont=bf,up,textfont=it,up]{caption} 
\usepackage{booktabs} 

\usepackage{lettrine} 

\usepackage{enumitem} 
\setlist[itemize]{noitemsep} 

\usepackage{abstract} 

\usepackage{titlesec} 
\renewcommand\thesection{\Roman{section}} 
\renewcommand\thesubsection{\roman{subsection}} 
\titleformat{\section}[block]{\large\scshape\centering}{\thesection.}{1em}{} 
\titleformat{\subsection}[block]{\large}{\thesubsection.}{1em}{} 

\usepackage{fancyhdr} 
\usepackage{titling} 

\usepackage{hyperref} 

\usepackage{verbatim} 

\usepackage{amstext} 

\usepackage{amsmath} 

\pretitle{\begin{center}\Large\bfseries} 
\posttitle{\end{center}} 

\def\bea{\begin{eqnarray}}
\def\eea{\end{eqnarray}}
\def\ben{\begin{equation}}
\def\een{\end{equation}}
\def\benu{\begin{enumerate}}
\def\enu{\end{enumerate}}

\def\bei{\begin{itemize}}
\def\eei{\end{itemize}}
\def\beit{\begin{itemize}}
\def\eit{\end{itemize}}
\def\benu{\begin{enumerate}}
\def\enu{\end{enumerate}}

\def\n{n}

\def\sss{\scriptscriptstyle\rm}

\def\1var{(\bx_1...\bx\N)}

\def\half{\frac{1}{2}}

\def\br{{\bf r}}

\def\bx{{\bf x}}

\def\x{_{\sss X}}
\def\c{_{\sss C}}

\def\cv{_{{\sss C},v}}

\def\s{_{\sss S}}
\def\xc{_{\sss XC}}

\def\N{_{\sss N}}
\def\H{_{\sss H}}
\def\trad{^{\rm trad}}
\def\inv{^{\rm inv}}

\def\HF{^{\rm HF}}

\def\PBE{^{\rm PBE}}

\def\ee{_{\rm ee}}

\def\sph_int{ {\int d^3 r}}

\def\rxn{_{\sss rxn}}
\def\srxn{_{\sss S,rxn}}
\def\sest{_{\sss S,est}}
\def\srxnest{_{\sss S,rxn,est}}
\def\crxn{_{\sss C,rxn}}
\def\cest{_{\sss C,est}}
\def\crxnest{_{\sss C,rxn,est}}
\def\eq{_{\sss e}}
\def\st{_{\sss s}}
\def\F{_{\sss F}}
\def\D{_{\sss D}}
\def\PBE{^{\sss PBE}}
\def\CC{^{\sss CC}}

\title{Measuring density-driven errors using Kohn-Sham inversion} 
\author{%
\textsc{Seungsoo Nam, Suhwan Song, and Eunji Sim}\thanks{esim@yonsei.ac.kr} \\ 
\normalsize Department of Chemistry, Yonsei University, 50 Yonsei-ro Seodaemun-gu, Seoul 03722, Korea \\ 
\textsc{Kieron Burke} \\ 
\normalsize Departments of Chemistry and of Physics, University of California, Irvine, CA 92697, USA \\  
}
\date{\today} 

\renewcommand{\maketitlehookd}{
\begin{abstract}
\noindent 
Kohn-Sham inversion, that is, the finding of the exact Kohn-Sham potential for a given density, is difficult in localized basis sets.
We study the precision and reliability of several inversion schemes, finding estimates of density-driven errors at a useful level of accuracy.  
In typical cases of substantial density-driven errors, HF-DFT is almost as accurate as DFT evaluated on CCSD(T) densities.
A simple approximation in practical HF-DFT also makes errors much smaller than the density-driven errors being calculated.
Two paradigm examples, stretched NaCl and the HO$\cdot$Cl$^-$ radical, illustrate just how accurate HF-DFT is.
\end{abstract}
}

\begin{document}

\maketitle

\sf

\section{Introduction}

Kohn-Sham (KS) density functional theory (DFT)\cite{KS65} is an extremely popular approach to electronic
structure problems, but the quality of the results depends on the quality of the
exchange-correlation (XC) approximation used.   Because the KS equations are solved
self-consistently, there are errors in both the self-consistent (SC-) energy and the SC-density.\cite{KSB13}  
In most KS calculations, the density errors contribute little to the overall
energy error.  However, in various generic situations, semilocal approximations to
XC make unusually large density errors (called density-driven errors) and the density
error contributes significantly to the resulting energy error.   In modern DFT parlance,
these are attributed to delocalization errors of the density.\cite{CMY08}

The theory behind density-corrected (DC) DFT explains the origin of such 
errors, when they are likely to be significant, and how they can usually be reduced by using a more accurate
density.\cite{KSB13, WNJK17, VSKS19}
The exact density-driven error is defined as the difference in energy when
the approximate functional is evaluated on its SC and exact densities.  
If the exact density was needed to perform a DC-DFT calculation, the procedure would be
impractical, as finding a highly accurate density is more costly than the DFT calculation
itself.  However, in practice, for many semilocal approximations applied to molecular
properties, it has been found that most density-driven errors can be greatly reduced
by use of the Hartree-Fock (HF) density instead of the exact density.   As the HF density
is of comparable cost to DFT, this leads to a very practical approach (HF-DFT) which
can be implemented very rapidly and costs no more than a typical DFT calculation.\cite{HF-DFT}
HF-DFT has been used to reduce density-driven errors for 
electron affinities,\cite{KSB11} potential energy curves,\cite{KSB14, KPSS15} 
spin gaps for coordination compound,\cite{SKSB18} and noncovalent interactions.\cite{KSEB18}

Given these successes of HF-DFT, we now ask: Can its underlying assumptions be tested?
The most important assumption is that, when density-driven errors are significant, in molecules, the HF density yields more accurate energies than the SC density.
 A lesser assumption is that, in practical HF-DFT calculations, the difference between HF and KS kinetic energies are ignored.
The answer is yes, by employing the well-established technique of KS inversion to
highly accurate densities, in order to extract exact density-driven errors and compare
with the HF-DFT procedure.
KS inversion is the process of finding an accurate KS potential
(and associated objects, such as KS kinetic energy, HOMO, etc.) from a given density.
From very early on,\cite{AP84} method-developers in DFT
have sought such exact information.\cite{VB94, GB96, UG94, BCDH94, BGHI05}
But most such inversions have been focussed on specific quantities such as eigenvalues,
which can be very sensitive to the details of the density.

Here, we apply standard KS inversion procedures with the sole focus of testing the 
assumptions underlying HF-DFT.  We use inversions to calculate density-driven errors
for typical systems in which HF-DFT has proven successful.  With two standard
methods, we explore both the dependence on the basis set and the guiding density
functional used (defined later).   
However, there are well-documented difficulties\cite{HIGB01, SSD06, HBY07, KN14} when such inversions are performed in finite localized basis sets.
We find that methods to overcome such difficulties, while imprecise, yield sufficient accuracy to answer the most basic questions about the density-driven error.
These methods, applied to HF and high-level $ab~initio$ densities in standard basis sets, produce sufficiently accurate
density-driven error estimates to usefully address such questions, i.e., their 
remaining errors are small relative to common density-driven errors.  We also
find that the approximation used in practical HF-DFT calculations, namely 
using the HF kinetic energy instead of the KS kinetic energy, typically leads
to changes of 1-2~kcal/mol, which is below a standard threshold for declaring
a density-driven error.\cite{SSB18}   

The paper is organized as follows. 
In Section II, we present backgrounds about wavefunctions and KS-DFT, KS inversion, and DC-DFT.  
Section III shows inversion results and gives some discussions about the uncertainty of the inversion, together with testing DC-DFT.
Finally, we deduce our conclusion from the discussions.


\section{Background}

\subsection{Wavefunctions and Kohn-Sham DFT}

Start from the variational principle for the exact ground-state energy
\ben
E_v=\min_\Psi \langle \Psi | \hat H | \Psi \rangle,
\een
where $\hat H$ is the $N$-electron Hamiltonian with one-body potential
$v(\br)$, and the search is over all
antisymmetric, normalized many-body wavefunctions $\Psi$. 
A HF calculation uses only a single Slater
determinant, denoted $\Phi$ (assuming for now no symmetry breaking):
\ben
E\HF_v=\min_\Phi \langle \Phi | \hat H | \Phi \rangle,
\een
where $\Phi\HF_v$ denotes the minimizer.  The traditional
definition of the correlation energy is then
\ben
E\trad\cv = E_v - E\HF_v, 
\een
and is non-positive because of the variational principle.

DFT replaces the central role of the one-body
potential with the ground-state density $\n(\br)$.  From the Hohenberg-Kohn
theorem,\cite{HK64} there is (at most) one $v(\br)$ which has a given density
as its ground state (for simplicity, we treat only non-degenerate
ground state here).
From the variational principle introduced in the Hohenberg-Kohn theorem, the ground state energy of system of $N$ electrons and external potential $v(\br)$ is\cite{L79,L83}
\begin{equation}
E_v = \min_{n \rightarrow N} \left( F[n] + \int v(\br)n(\br) d\br \right) ,
\label{eq:Etot_HK}
\end{equation}
where $F[n]$ is the universal part of the
Hohenberg-Kohn functional, defined as
\begin{equation}
F[n] = \min_{\Psi \rightarrow n} \langle \Psi|\hat{T}+\hat{V}\ee|\Psi \rangle ,
\label{eq:F_HK}
\end{equation}
where $\hat{T}$ is the kinetic energy operator, $\hat{V}\ee$ is the
electron repulsion operator, and the 
minimization is over all antisymmetric wavefunctions that integrate to
density $n(\br)$. Denote the minimizer by $\Psi[\n]$.
The further ansatz of the KS scheme is that there exists a
local multiplicative potential, $v\s(\br)$, whose ground-state density
for non-interacting fermions matches the interacting one.
The total energy in terms of KS quantities is then
\begin{equation}
E_v=\min_{\n}\left(T\s[n] + \int v(\br)n(\br) d\br + E\H[n] + E\xc [n]\right),
\label{eq:Etot_KS}
\end{equation}
where $T\s$ is the kinetic energy of the KS electrons,
$E\H[n]$ is the Hartree energy, and $E\xc [n]$ is the XC energy. 
The KS wavefunction is $\Phi\s[\n]$, which we take here to be a single Slater determinant,
as is typical.

We highlight some subtle points that will be important in what follows.
The quantum-mechanical operators are known, so each energy components
has an obvious functional dependence on the wavefunction, such as
\ben
T[\Psi] =-\half\sum_{i=1}^N \langle \Psi | \nabla^2_i | \Psi \rangle
\een
in atomic units.   
For any Slater determinant $\Phi=\{\chi\}_N$ of $N$ orbitals $\chi_i(\bx)$ of space-spin coordinate $\bx=(\br, \sigma)$ and $\int d\bx = \sum_{\sigma}\int d\br$,
\ben
T[\Phi] = \half \sum_{i=1}^N \int d\bx |\nabla \chi_i (\bx) |^2.
\een
Density functionals are then defined via the minimizing wavefunctions.
The KS kinetic energy is found by setting $\hat{V}\ee=0$ in Eq.~\ref{eq:F_HK}
\ben
T\s[n]=\min_{\Psi \rightarrow n} \langle \Psi | \hat{T} | \Psi \rangle = T[\Phi\s [n]] ,
\een
where $\Phi\s [n]$ is minimizer.
The exact kinetic energy is 
\ben
T[n]= T[\Psi [n] ] .
\een
These two differ by the correlation kinetic energy:
\ben
T\c[\n]=T[\n]-T\s[\n],
\een
which must be non-negative, as $T\s$ is the minimizer of $\hat T$ for the given 
density. 
Analogously, the exchange energy of $N$ orbitals is:
\begin{equation}
E\x[\Phi] = -\frac{1}{2}\sum_{i,j}^{N}\iint d\bx d\bx' \frac{\chi_i^*(\bx) \chi_j^*(\bx') \chi_j(\bx) \chi_i(\bx')}{|\br-\br'|},
\label{eq:Ex}
\end{equation}
which yields the exact exchange density functional in DFT:
\ben
E\x[\n] = E\x[\Phi\s[\n]].
\een
The DFT definition of the correlation energy is then 
\begin{equation}
\begin{split}
E\c[\n] &= \langle \Psi[\n] | \hat H | \Psi[\n] \rangle - \langle \Phi\s[\n] | \hat H | \Phi\s[\n] \rangle  \\
&= T\c[\n]+U\c[\n],
\end{split}
\end{equation}
\noindent where the potential contribution to correlation is
\ben
U\c[\n]=V\ee[\n]-E\H[\n]-E\x[\n].
\een
For weakly correlated systems, such as atoms or most molecules,
 $T\c$ is only slightly less than $|E\c|$, so that $E\c+T\c$
which is non-positive, is much smaller in magnitude than either.
For example, for the He atom, $E\c$ is -42~mH, $T\c$ is 36~mH, and their
sum is -6~mH.\cite{HU97}

There are subtle differences between DFT and wavefunction theory.\cite{GP96}
Since the HF Slater determinant minimizes
$\hat H$ over all Slater determinants whereas the KS Slater determinant
is restricted to orbitals coming from a single multiplicative potential,
the definition of correlation energy differs:
\ben
\Delta E\cv = E\cv\trad - E\c[\n_v],
\een
so that
\ben
\Delta E\cv = \langle \Phi\s[\n_v] | \hat H | \Phi\s[\n_v] \rangle -
\langle \Phi\HF_v | \hat H | \Phi\HF_v \rangle
\een
must be non-negative (although only very slightly, 0.1~mH for He).\cite{GP96}
A larger difference comes from the difference between the KS and HF
Slater determinants, even when they refer to the same density.  Define
\ben
\Delta T\s\HF[\n]= T[\Phi\HF_{v[n]}] - T\s[\n] \geq 0,
\label{eq:dTsHF}
\een
where $\Phi\HF_{v[n]}$ is the HF Slater determinant of $v[\n](\br)$, the one-body potential whose exact density is $n(\br)$.
We call $\Delta T\s\HF[\n]$ the excess non-interacting HF kinetic energy.
(In principle, this is found by adjusting $v(\br)$ until a HF calculation
yields $\n(\br)$ as its density.)
This must be non-negative, as the KS kinetic energy is the minimizer, and can
be several mH for mid-size atoms.\cite{GE95}
Moreover, as $E[\Phi\HF_{v[\n]}] \leq E[\Phi\s[\n]]$, and both Hartree and one-body terms cancel,
\ben
\Delta E\x\HF[n] \leq -\Delta T\s\HF[\n],
\label{eq:ExTs}
\een
where 
\ben
\Delta E\x\HF[\n]= E\x[\Phi\HF_{v[\n]}] - E\x[\n].
\een

In the special case of atoms, the virial theorem guarantees that the
total energy is exactly the negative of the kinetic energy for any minimized
calculation, either exact, HF, or DFT with some XC functional.  
This implies
\ben
\Delta E\x\HF[n] = -2\Delta T\s\HF[n]~~~{\rm (atoms)}
\label{eq:virial}
\een
exactly.

\subsection{Kohn-Sham Inversion}

The problem of finding accurate KS potentials for given
densities has been studied almost as long as KS-DFT has been used.\cite{AP84}
There are now many algorithms in existence and use.\cite{G92, WP93, VB94, ZMP94, WY03, RS12, ZC18, FAB18, KZG19}
Some use just the density itself (a pure KS inversion) whereas others are focused
on the most relevant case, i.e., densities generated by more
accurate and controllable wavefunction calculation, in which case
more information is available and can be used.\cite{RKS15, CAS15, CS16, ORS17, OS18}
Here we use two pure KS inversion schemes, Zhao-Morrison-Parr (ZMP)\cite{ZMP94} and Wu-Yang (WY).\cite{WY03}
We always assume the target density is pure-state non-interacting $v$-representable.

The solvers typically work by iteration.  A guess for the desired KS
potential is made, the KS orbitals are generated, the density is
calculated, and the guessed potential is updated according to some
algorithms.  There are several relevant convergence criteria.
The first is the choice of basis set for the inversion algorithm (note that
this is independent of the basis set used to generate the target density).
Second, there are often guiding functions for the guess.  
Because typical XC approximations have incorrect behavior of $v\s(\br)$ far from
nuclei, one often uses the Fermi-Amaldi (FA) potential,\cite{FA34} which builds in the
correct behavior.
In the case of the ZMP procedure, there is a penalty function for errors in the
density, which is multiplied by a dimensionless parameter $\lambda$.
As $\lambda\to\infty$, the procedure converges to the target density,
but it can become unstable for too large values of $\lambda$.   

Traditionally, such inversions are performed on accurate densities 
in order to gain insight into the exact KS quantities.  The paradigmatic
example is the extraordinary usefulness of the atomic KS potentials
produced by Umrigar and co-workers.\cite{UG94, BCDH94}
Knowledge of the positions of the exact KS eigenvalues has been 
invaluable in tests of time-dependent DFT.\cite{SUG98}
But the KS orbital energies are extremely
sensitive to details of the potential, but ground-state energy differences 
(e.g., reaction energies) are not.
Below we introduce just those quantities that are relevant to DC-DFT as criteria
for sufficient convergence of the KS inversion.
For a given inversion recipe, $\n\inv(\br)$ is a functional of the input
or target density, $\n(\br)$, as are all the KS inversion orbitals and eigenvalues.
For any energy functional of the density, we define the
inversion error as
\ben
\Delta A\inv[\n]=A[\n\inv[\n]]-A[\n],
\label{eq:inv_err}
\een
and we want this error to be sufficiently small so as not to obscure the 
reaction energies we wish to calculate.

The inversion error in a finite basis comes from several sources.
First, a density from a multi-determinantial wavefunction, that is, a correlated density, in a given finite atomic basis set typically cannot be exactly expressed as a KS density in that basis, and a more extensive basis set is required (see \cite{MPBN17} for details).
This problem only occurs when inverting a correlated target density.
Although the basis set of the target density and inversion need not be the same in principle,
we usually use same basis set because ZMP and WY typically work with density matrices.
Second, one cannot provide infinite flexibility to the KS potential in practice.
In the case of ZMP, potential flexibility is limited to the size of the atomic orbital basis set.
On the other hand, WY introduces a potential basis, which allows one to increase the flexibility of the potential by increasing the size of that basis.
However, increased flexibility of the KS potential in WY may produce orbitals that are very close to the HF orbitals of the same density.\cite{SSD06}
This makes the actual computation of Eq.~\ref{eq:inv_err} for the KS kinetic energy (i.e., $A=T\s$) impossible because when $\n\inv[\n]$ approaches to $\n$, then simultaneously $T\s$ incorrectly approaches to $T[\Phi\HF_{v[\n]}]$.
Nevertheless, in the later section, we will show that this ambiguity is sufficiently small as to not invalidate our results.

\subsection{Density Corrected DFT}

DC-DFT claims that, under well-understood conditions, the
SC density in an approximate DFT calculation can contribute
significantly to the error, and that such error can usually be reduced by
use of a more accurate density.
The conventional measure for DFT error in energy is
\begin{equation}
\Delta E  = \tilde{E}[\tilde{\n}] - E[\n],
\label{eq:TE}
\end{equation}
where $E$ is the exact energy functional (of Eq.~\ref{eq:Etot_KS} in KS-DFT), and tilde denotes an approximation. 
One can define a functional error that comes from the approximate $\tilde{E}$ only, by
\begin{equation}
\Delta E\F = \tilde E[\n] - E[\n] = \tilde E\xc[\n]-E\xc[\n],
\label{eq:FE}
\end{equation}
where the last equality holds in a KS calculation.
The rest of the error comes from the $\tilde{\n}(\br)$ in the given energy functional;
\begin{equation}
\Delta E\D = \Delta E - \Delta E_{\sss F} = \tilde E [\tilde \n] - \tilde E [\n],
\label{eq:DDE}
\end{equation}
and is called the density-driven error.\cite{KSB13,KSB14}

In DC-DFT, in principle, one should apply the approximate functional to the exact density for DFT calculations
whose density-driven errors are significant (about 2~kcal/mol for small molecules.\cite{SSB18}).  
By eliminating the density-driven error, the energy usually improves significantly.\cite{KSB11, KSB14, KPSS15, SKSB18, KSEB18}
In practice, calculating highly accurate densities, such as from CCSD(T), is similar to or more expensive than simply running CCSD(T) to find energies.
For molecular calculations, HF-DFT often suffices to yield significantly improved energetics when density-driven errors are large,\cite{KSB11, KSB14, KPSS15, SKSB18, KSEB18} with little or no additional cost relative to the SC-DFT calculation.
However, HF-DFT uses the HF orbitals, simply swapping the 
HF exchange for the approximate DFT XC, evaluated on the HF orbitals.
This procedure ignores $\Delta T\HF\s$, the difference between HF and KS kinetic energies.


\section{Results and Discussion}

Our aim is to test the WY and ZMP KS inversion schemes for use in validating the assumptions underlying HF-DFT.
Thus inversion errors in energies must be smaller than the density-driven errors that are (presumably) being eliminated by the HF-DFT procedure.
In this section, we perform inversions targeting KS, HF, and correlated density, and check the accuracy and precision of the inversion.
As our prototypical choices, we consider the NaCl molecule, both at equilibrium
($R\eq=2.4~\AA$) and when stretched ($R\st=4.5~\AA$).  
At equilibrium, most calculations with standard functionals are normal (DFT error has negligible density-driven contribution), while most are abnormal when stretched.\cite{KSB13, KSB14, KPSS15}
Our default (standard) functional is PBE, and our default (standard) basis set is aug-cc-pVTZ.
We introduce shorthand notations for the potential basis sets (PBS) for WY; X, CX, uCX, and ACX, stand for cc-pVXZ, cc-pCVXZ, unc-cc-pCVXZ, and aug-cc-pCVXZ, respectively, where X(=D, T, Q, 5) is the cardinal number of the PBS.

\subsection{Approximate KS target density}

\begin{table}[h!]
\footnotesize
\centering
\begin{tabular}{c c r r r r r r}
\hline \hline
\multicolumn{2}{c}{Functional}    & \multicolumn{2}{c}{$\Delta E\PBE$} & \multicolumn{2}{c}{$\Delta E\PBE\rxn$} & \multicolumn{2}{c}{$\Delta T\srxn$} \\
\multicolumn{2}{c}{Geometry}      & $R\eq$           & $R\st$          & $R\eq$             & $R\st$            & $R\eq$           & $R\st$           \\ 
\hline
guide                 & $\lambda$ & \multicolumn{6}{c}{ZMP}                                                                                           \\
\cline{3-8}
\multirow{4}{*}{FA}   & 64        & 11.82            & 11.61           & 3.22               & 3.01              & -534             & -524             \\
                      & 128       & 7.08             & 6.90            & 1.62               & 1.44              & -309             & -300             \\
                      & 256       & 4.08             & 3.98            & 0.83               & 0.72              & -178             & -171             \\
                      & 512       & 2.06             &                 & 0.37               &                   & -101             &                  \\
\cline{3-8}
\multirow{4}{*}{SVWN} & 64        & 0.40             & 0.39            & 0.01               & 0.00              & -4.88            & -3.96            \\
                      & 128       & 0.27             & 0.26            & 0.01               & 0.00              & -4.46            & -3.65            \\
                      & 256       & 0.17             & 0.16            & 0.01               & 0.00              & -3.68            & -2.70            \\
                      & 512       & 0.10             &                 & 0.01               &                   & -2.67            &                  \\
\cline{3-8}
\multirow{4}{*}{BLYP} & 64        & 0.04             & 0.04            & 0.00               & 0.00              & -0.95            & -0.03            \\
                      & 128       & 0.02             & 0.02            & 0.00               & 0.00              & -1.28            & -0.22            \\
                      & 256       & 0.01             & 0.01            & 0.00               & 0.00              & -1.15            & -0.17            \\
                      & 512       & 0.01             &                 & 0.00               &                   & -0.83            &                  \\
\hline
guide                 & PBS       & \multicolumn{6}{c}{WY}                                                                                            \\
\cline{3-8}
\multirow{4}{*}{FA}   & D         & 1.81             & 1.82            & -0.44              & -0.43             & 3.08             & 10.12            \\
                      & T         & 0.53             & 0.34            & 0.06               & -0.13             & -3.01            & 0.40             \\
                      & CT        & 0.03             & 0.02            & 0.01               & 0.00              & 0.28             & 0.25             \\
                      & CQ        & 0.00             & 0.00            & 0.00               & 0.00              & 0.10             & 0.16             \\
\cline{3-8}
\multirow{4}{*}{SVWN} & D         & 0.34             & 0.49            & -0.03              & 0.11              & -3.60            & -0.48            \\
                      & T         & 0.12             & 0.09            & 0.03               & 0.00              & -3.18            & -0.89            \\
                      & CT        & 0.01             & 0.00            & 0.00               & 0.00              & 0.09             & 0.25             \\
                      & CQ        & 0.00             & 0.00            & 0.00               & 0.00              & -0.03            & 0.21             \\
\cline{3-8}
\multirow{4}{*}{BLYP} & D         & 0.06             & 0.12            & -0.02              & 0.04              & -0.60            & 1.04             \\
                      & T         & 0.02             & 0.02            & 0.00               & 0.00              & -1.31            & -0.25            \\
                      & CT        & 0.00             & 0.00            & 0.00               & 0.00              & 0.36             & 0.26             \\
                      & CQ        & 0.00             & 0.00            & 0.00               & 0.00              & 0.04             & 0.22             \\
\hline \hline
\end{tabular}
\caption{
Inversion errors on a KS density for total molecular energy ($\Delta E$) reaction energy ($E\rxn$=$E$(mol)-$E$(atoms)), and KS kinetic energy for reaction, for PBE calculations on NaCl, in aug-cc-pVTZ basis. Here, $R\eq$=2.4~$\AA$, $R\st$=4.5~$\AA$ and blank cells denote inversions are not converged.
All energies are in mH.
}
\label{table:inv2pbe}
\end{table}

A simple consistency check is to take the density from a standard DFT calculation 
and run inversions to see how accurately we recover the KS energetic components, for which we have 'exact' answers from the original calculation.
To avoid trivial solutions, we used guiding potentials that are not used for SC calculation.
In the case of a KS target density, one can easily calculate Eq.~\ref{eq:inv_err} for the total energy
because $T\s$ is known from the SC-KS calculation.
Table \ref{table:inv2pbe} shows results for NaCl with PBE and its inverted densities.  
Several important lessons can be drawn from these results.
First, errors in this inversion can be driven down to the microhartree range.
Second, errors are typically reduced by tightening the convergence parameters,
such as larger PBS, larger values of $\lambda$, or using guiding functionals
that are close to the original functional that generated the density.
Third, when convergence is an issue, total energy converges much faster than energy components, and 
reaction energies converge much faster than individual energies. 
Fourth, the FA guiding functional converges most slowly here, presumably
because the PBE target density was generated from an XC functional yielding a
different (and incorrect) asymptotic behavior.
Nevertheless, for ZMP/FA, $\lambda = 512$ yields sufficiently accurate reaction energies, (subscript ${\sss rxn}$ hereafter)
so we chose to use $ \lambda = 512 $ as our default.
In the case of WY, since the accuracy of $\Delta T\srxn$ is greatly improved when we increase PBS from T to CT, we chose CT, i.e., cc-pCVTZ, as our default potential basis.

\subsection{Hartree-Fock target density}

\begin{table}[h!]
\small
\centering
\begin{tabular}{c c c r r r r}
\hline \hline
\multicolumn{2}{c}{Functional} & \multicolumn{2}{c}{$\Delta T\sest$} & \multicolumn{2}{c}{$\Delta T\srxnest$} \\
\multicolumn{2}{c}{geometry}  & $R\eq$ & $R\st $  & $R\eq $  & $R\st $     \\
\hline
guide                  & $\lambda$     & \multicolumn{4}{c}{ZMP}                                \\
\cline{3-6}
FA                     & 512   & 2.81        & 2.38       & 0.37         & -0.05        \\
BLYP                   & 512   & 2.36        & 1.95       & 0.22         & -0.18        \\
\hline
guide                  & PBS   & \multicolumn{4}{c}{WY}                                 \\
\cline{3-6}
\multirow{2}{*}{FA}    & CT    & 0.74        & 0.46       & -0.08        & -0.37        \\
                       & uC5   & 0.26        & 0.24       & 0.18         & 0.16         \\
BLYP                   & CT    & 0.77        & 0.48       & -0.07        & -0.36       \\
\hline \hline
\end{tabular}
\caption{
Estimated $\Delta T\HF\s$ values (Eq.~\ref{eq:dTHF_est}) for equilibrium ($R\eq=2.4~\AA$) and stretched ($R\st=4.5~\AA$) geometry of NaCl, and for corresponding reaction energies.
All energies are in mH.
}
\label{table:inv2HF}
\end{table}

\begin{table}[h!]
\small
\centering
\begin{tabular}{c c c r r r r}
\hline \hline
\multicolumn{2}{c}{Functional} & \multicolumn{2}{c}{$\Delta E\HF_{\sss X,est}$} & \multicolumn{2}{c}{$\Delta E\HF_{\sss X,rxn,est}$} \\
\multicolumn{2}{c}{geometry}  & $R\eq$ & $R\st $  & $R\eq $  & $R\st $     \\
\hline
guide                  & $\lambda$     & \multicolumn{4}{c}{ZMP}                                \\
\cline{3-6}
FA            & 512            & -7.60      & -6.69     & -1.29       & -0.38       \\
BLYP          & 512            & -3.70      & -2.94     & -0.70       & 0.06        \\
\hline
guide                  & PBS   & \multicolumn{4}{c}{WY}                                 \\
\cline{3-6}
FA            & CT             & -1.70      & -1.13     & 0.30        & 0.88        \\
FA            & uC5            & -0.56      & -0.53     & -0.39       & -0.36       \\
BLYP          & CT             & -1.68      & -1.04     & 0.14        & 0.78       \\
\hline \hline
\end{tabular}
\caption{
Estimated $\Delta E\x\HF$ values for equilibrium ($R\eq=2.4~\AA$) and stretched ($R\st=4.5~\AA$) geometry of NaCl, and for corresponding reaction energies.
All energies are in mH.
}
\label{table:inv2HF_Ex}
\end{table}

Our first non-trivial task is to find $\Delta T\s\HF$, the contribution ignored in a typical HF-DFT calculation.
A HF-DFT calculation first runs a HF calculation, then replaces the exchange term with the XC of KS-DFT.
For any approximate XC functional, define
\ben
\tilde{V}_{\sss exp}[\n] = \int d\br n(\br) v(\br) +E\H[\n]+\tilde{E}\xc [n],
\een
the contributions to the energy that are known explicitly as functionals of the density. 
Then,
\ben
\tilde E^{\sss HF-DFT} = T[\Phi\HF_v] +\tilde{V}_{\sss exp}[\n_v\HF].
\label{eq:HF-DFT}
\een
However, DFT energies on HF densities, are defined as:
\ben
\tilde E[\n\HF] = T\s[n\HF] + \tilde{V}_{\sss exp}[\n\HF].
\label{eq:DFT[HF]}
\een
Subtracting Eq.~\ref{eq:DFT[HF]} from both sides of Eq.~\ref{eq:HF-DFT} when $\n\HF=\n\HF_v$ yields
\ben
\tilde{E}^{\sss HF-DFT}_v -\tilde{E}[n\HF_v] = T[\Phi\HF_v]-T\s[n\HF_v] = \Delta T\HF\s[n\HF_v].
\label{eq:dTHF}
\een
However, due to the uncertainty of the inversion, we can not calculate  $\tilde E[\n\HF]$ nor $T\s[n\HF]$ exactly.
Instead, we calculate $\tilde E[n\inv[n\HF]]$ or $T\s[n\inv[n\HF]]$ via approximate inversion of the HF density.
Because the total energy is much less sensitive to small changes of the density (here, inversion error) than individual  energy components such as the kinetic energy, for the estimation of $\Delta T\s\HF$, we use
\ben
\Delta T\HF\sest = \tilde E^{\sss HF-DFT}_v-\tilde E[\n\inv[n\HF_v]],
\label{eq:dTHF_est}
\een
where the subscript ${\sss est}$ represents the estimated value.

Table~\ref{table:inv2HF} reports $\Delta T\HF\sest$ values (Eq.~\ref{eq:dTHF_est}) for 
Na and Cl atoms and reaction (atomization) energies of NaCl.
Even with our standard protocol, the $\Delta T\s\HF$ of ZMP and WY in molecular energy varies by up to 2~mH.
But reaction energies are far less sensitive, and here variations are negligible.
As the reaction energies of HF-PBE are -148 and -62~mH for equilibrium and stretched geometry, respectively, these variations are less than 1\% of the reaction energy.
Therefore, although obtaining a precise $\Delta T\s\HF$ is not possible with our methodology, estimations can be made precisely enough (within about $\pm$0.5~mH) for reaction energies to be useful in testing HF-DFT.

In a WY inversion, $\Delta T\HF\sest$ becomes very small when PBS is very large, for example, uC5.
In the reaction energy calculation of NaCl, this does not cause a severe problem because $\Delta T\HF\s$ itself is very small. 
However, when $\Delta T\HF\s$ is large, then increasing the size of PBS will eventually cause $\Delta T\HF\sest$ to vanish incorrectly.
We find that a PBS with the same level of $\zeta$ plus tight core functions is balanced (giving an accurate density but not an unphysically small $\Delta T\HF\sest$) with an atomic orbital basis.
Also, $\Delta T\HF\sest$ using different guiding potentials using CT PBS varies less than 0.03~mH,
indicating that CT is flexible enough to cover the differences in various guiding potentials.
We thus report values of $\Delta T\HF\s$ for reaction energies only, using WY and assuming uncertainties of $\pm$0.5~mH.

To take advantage of error cancellations for $\Delta E\x\HF$, we define
\ben
\Delta E\HF_{\sss X,est}
=E_v\HF- \langle \Phi\s[n\inv[n\HF_v]] | \hat{H} | \Phi\s[n\inv[n\HF_v]] \rangle -\Delta T\HF\sest,
\een
which are shown in Table~\ref{table:inv2HF_Ex}.
For each inversion, $\Delta E\HF_{\sss X,est} < 0$ for total energies. (Eqs.~\ref{eq:dTsHF} and \ref{eq:ExTs})
Also, by compairing Tables~\ref{table:inv2HF} and \ref{table:inv2HF_Ex}, typically $\Delta E\HF_{\sss X,est} \approx -2\Delta T\HF\sest$, both for total and reaction energies, as expected from Eq.~\ref{eq:virial}.
Deviation from atomic results should be small because NaCl is a weakly correlated system.
In the case of ZMP, $\Delta E_{\sss X,est}\HF \approx -2\Delta T\HF\sest$ is less clear, which shows the limits of the inversion accuracies on this scale of energies.

\subsection{Correlated target density}

\begin{table}[h!]
\small
\centering
\begin{tabular}{c c c r r r r}
\hline \hline
\multicolumn{2}{c}{Functional} & \multicolumn{2}{c}{$T\cest$} & \multicolumn{2}{c}{$T\crxnest$} \\
\multicolumn{2}{c}{geometry}  &$R\eq $ &$R\st $  & $R\eq $  & $R\st $   \\
\hline
guide                  & $\lambda$     & \multicolumn{4}{c}{ZMP}                                \\
\cline{3-6}
FA                     & 512   & 193.10        & 189.01      & 21.77 & 17.68        \\
BLYP                   & 512   & 193.12        & 189.07       & 21.80         & 17.74        \\
\hline
guide                  & PBS   & \multicolumn{4}{c}{WY}                                 \\
\cline{3-6}
FA                     & CT   & 191.94        &187.77       & 21.85        & 17.68        \\
BLYP                   & CT   & 191.97        & 187.78       & 21.87         & 17.68        \\
\hline \hline
\end{tabular}
\caption{
Estimated $T\c$ values for equilibrium ($R\eq=2.4~\AA$) and stretched ($R\st=4.5~\AA$) NaCl, and for corresponding reaction energies.
Estimations were made using Eq.~\ref{eq:dTc_est}. 
All energies are in mH.
}
\label{table:inv2CC}
\end{table}

Now we consider the inversion when targeting an electron density from a correlated CCSD wavefunction ($\n\CC$). 
To check the quality of such an inversion, we extract $T\c$, the kinetic correlation energy, as accurately as practical. 
Just as for HF, to take advantage of error cancellations, we define, analogous to Eq.~\ref{eq:HF-DFT},
\ben
\tilde{E}_v^{\sss TC-DFT}  = T[\Psi\CC_v] +\tilde{V}_{\sss exp}[\n_v\CC],
\een
where $T$ is the value from the CCSD calculation.
We can then estimate $T\c$ analogously to Eq.~\ref{eq:dTHF_est},
\ben
T\cest = \tilde{E}_v^{\sss TC-DFT}  - \tilde{E}[\n\inv[n_v\CC]].
\label{eq:dTc_est}
\een
Similar to Table~\ref{table:inv2HF}, standard ZMP and WY give different estimates of $T\c$ (by approximately 1.2~mH), but this variation is much smaller in reaction energies (approximately 0.1~mH). 

From Tables~\ref{table:inv2HF} and ~\ref{table:inv2CC}, it seems not possible to obtain exact $\Delta T\HF\s$ or $T\c$ values due to the fundamental limitations of the inversion methods.
However, the two different inversion algorithms, ZMP and WY, yield consistent estimates for $\Delta T\HF\srxn$ and $T\crxn$.
Therefore, we expect inversion can provide $\tilde E[\n]$ with minor uncertainty, allowing density-driven error estimates with a useful level of accuracy.
Since the deviation of the $\Delta T\HF\srxnest$ in Table~\ref{table:inv2HF} was up to 0.5~mH, we expect that the exact inversion results to be within $\pm$0.5~kcal/mol when calculated under standard inversion conditions. (Note that 1~mH < 1~kcal/mol.)
In the next subsection, as a practical application of the inversion, we will present the entire dissociation curves of NaCl and OH$\cdot$Cl$^-$.

\subsection{Testing HF-DFT}

\begin{figure} [h!]
\centering
\includegraphics[width=1.0\columnwidth]{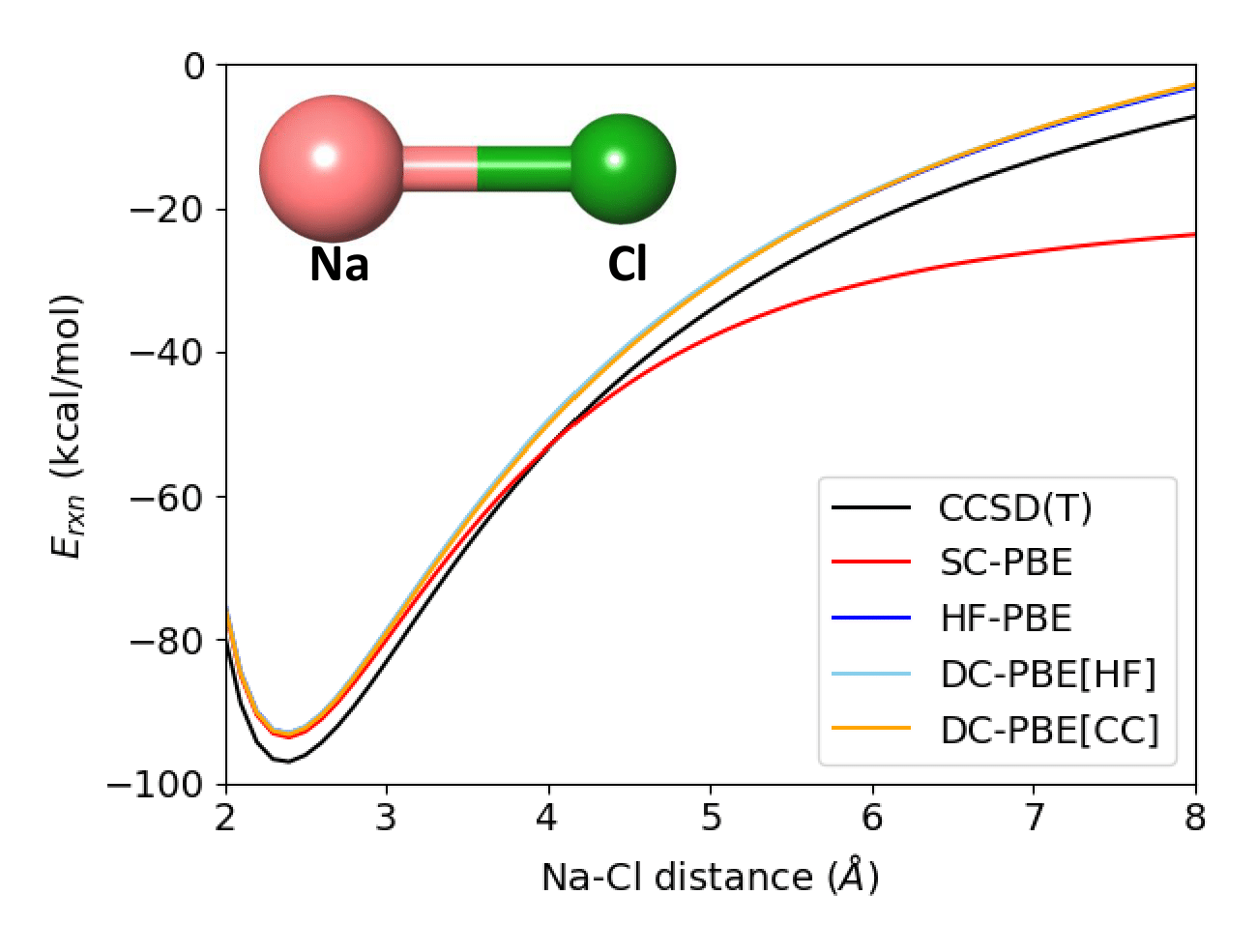}
\caption{
Dissociation curve of NaCl molecule using CCSD(T), SC-PBE, HF-PBE, and DC-PBE using HF or CCSD density, using WY/FA/CT inversion.
HF-PBE and DC-PBE[HF and CC] are indistinguishable on this scale.
The criterion for the inversion imprecision, $\pm$0.5~kcal/mol, is similar to the scale of the thickness of the lines.
}
\label{fgr:NaClpes}
\end{figure}

Previously, it has been argued that the poor description of SC-DFT calculations for dissociation limits of hetero-diatomic molecules was due to density-driven error, and HF-DFT reduces that error.\cite{KSB13, KPSS15}
Here we test the argument by quantitatively decomposing the DFT (here, PBE) error using a highly accurate CCSD density as a benchmark.
(CCSD(T) density shows no meaningful differences, see Table S10 in supporting information.)
In these calculations, all inversions were performed using the WY algorithm, FA guiding potential, and CT potential basis (WY/FA/CT).
Figure~\ref{fgr:NaClpes} presents dissociation curves of NaCl molecule.

While SC-PBE does not account for the correct dissociation behavior, making a huge well-known error in the dissociation limit,\cite{RPCV06} DC-PBE with the inversion of the HF or CCSD densities (DC-PBE[HF] and DC-PBE[CC], respectively) correctly capture the dissociation limit, although they are slightly above CCSD(T).
Most importantly, HF-DFT and either of these curves are indistinguishable, showing that HF-DFT differs negligibly from PBE energies evaluated on (essentially) exact densities.
This validates the use of HF-DFT as a practical approximation to DC-DFT.
Although we have defined density-driven error only for SC-DFT calculations, the density-driven error can be defined for any approximate XC energy for non-SC-DFT by replacing $\tilde n$ in Eq.~\ref{eq:DDE} to any non-SC density.\cite{VSKS19}
By doing so, we can calculate the density-driven error of $n\HF$.

\begin{figure} [h!]
\centering
\includegraphics[width=1.0\columnwidth]{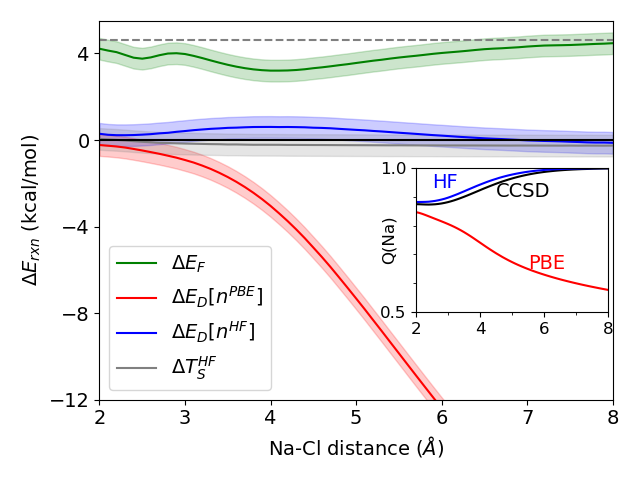}
\caption{
Functional error $\Delta E\F$ and density-driven error $\Delta E\D$ of $n\PBE$ and $n\HF$, and $\Delta T\HF\s$ of NaCl dissociation curve. 
Curves are drawn with $\pm$0.5~kcal/mol bands to represent the uncertainty of the inversion.
$\Delta T\HF\s$ is almost indistinguishable from zero.
Gray dashed horizontal line (4.6~kcal/mol) represents $\Delta E\F$ of the reaction Na + Cl $\to$ Na$^+$ + Cl$^-$.
Inset shows the IAO population of the Na atom.
}
\label{fgr:NaClerr}
\end{figure}

To further study the differences between these curves, Figure~\ref{fgr:NaClerr} shows the small differences between curves in Figure~\ref{fgr:NaClpes}.
We write 
\ben
\Delta E_{{\sss F},v}=\Delta E_v^{\sss HF-DFT} - \Delta E\D[n_v\HF] -\Delta T\s\HF[n_v\HF]
\een
by combining Eqs.~\ref{eq:FE}, \ref{eq:DDE}, and \ref{eq:dTHF}, where $\Delta E\D[n\HF]=\tilde{E}[n\HF]-\tilde{E}[n]$ is the PBE density-driven error of the HF density.
Curves are drawn with $\pm$0.5~kcal/mol bands to represent the uncertainty of the inversion.
$\Delta E\F$ in Figure~\ref{fgr:NaClerr} is almost constant regardless of the geometry.
Thus, on the scale of the PBE density-driven errors, our imperfect inversions definitely show that the functional error estimated by HF-DFT barely differs from the true value.

On the other hand, $\Delta E\D[n\PBE]$ grows strongly with Na-Cl distance, directly showing density delocalization error of PBE.
We observed almost zero $\Delta E\D[n\HF]$ for any geometry  in Figure~\ref{fgr:NaClerr}.
The behavior of $\Delta E\D$ can also be understood from population analysis.
Here, we used Mulliken population analysis using intrinsic atomic orbitals (IAO)\cite{K13} constructed from KS orbitals of either PBE or an inversion (HF or CCSD).
Note that IAO cannot be constructed directly from a correlated wavefunction, which requires KS inversion.
The results are shown in the inset of Figure~\ref{fgr:NaClerr}.
At Na-Cl distance 2~$\AA$, the population difference between CCSD and PBE is 0.03, which has almost no effect on $\Delta E\D[n\PBE]$.
The population difference between HF and CCSD is maximum near Na-Cl distance is 4~$\AA$.
This difference is reflected in the error curve, where $\Delta E\D[n\HF]$ becomes slightly positive at that geometry.
$\Delta T\HF\s$ is negligibly small everywhere (-0.08 and -0.37~mH for Na-Cl distance 2.4 and 4.5~$\AA$, see Table~\ref{table:inv2HF}).
The population of Na atom drops to zero after 8~$\AA$, where the triplet state becomes ground state (not shown).\cite{KPSS15}

Finally, we point out that the functional error in the reaction Na+Cl$\rightarrow$Na$^+$+Cl$^-$ (marked by the dashed horizontal line in Figure~\ref{fgr:NaClerr}) dominates the $\Delta E\F$ in Figure~\ref{fgr:NaClerr}.
Thus, when measured relative to this limit, the error in a PBE curve evaluated on CC densities, is $<$ 2~kcal/mol everywhere!

\begin{figure} [h!]
\centering
\includegraphics[width=1.0\columnwidth]{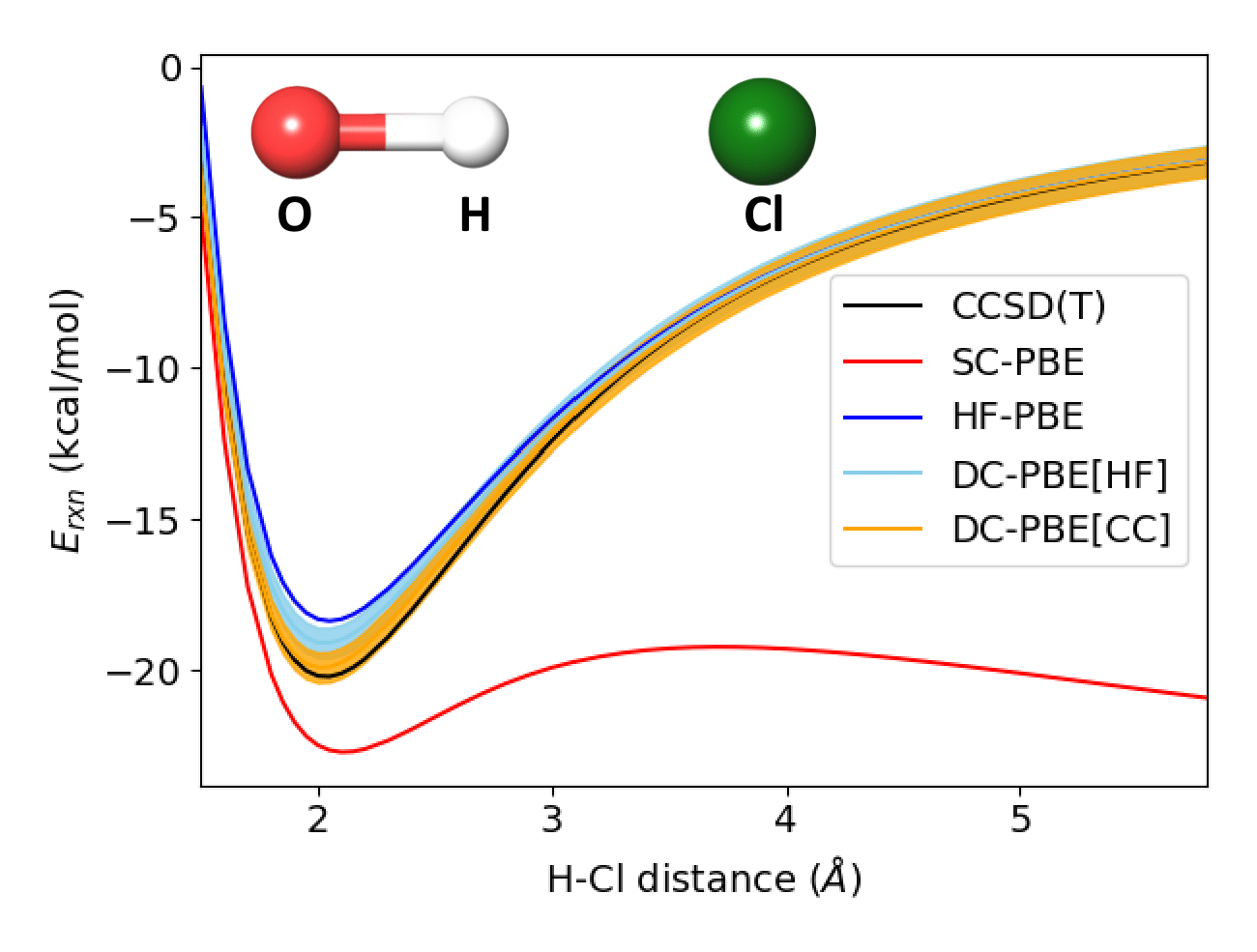}
\caption{
Dissociation curve of OH$\cdot$Cl$^-$ complex using CCSD(T), SC-PBE, HF-PBE, and DC-PBE with HF and CCSD density, using inversion.
DC-PBE curves are drawn with $\pm$0.5~kcal/mol bands to represent the uncertainty of the inversion.
All inversions were performed with WY/FA/ACT.
}
\label{fgr:HOClpes}
\end{figure}

As another example from a previous successful application of HF-DFT,\cite{KPSS15}
we also analyzed the potential energy curve of the linear HO$\cdot$Cl$^-$ complex (O-H distance fixed to $1~\AA$) in Figure~\ref{fgr:HOClpes}.
Here, we used augmented PBS (ACT) for WY, due to the WY convergence issue for H-Cl beyond $4~\AA$.
SC-PBE shows a significant deviation from CCSD(T), not only in the stretched geometry but even in the equilibrium geometry.
DC-PBE[CC] curve almost coincides with the reference CCSD(T), showing that $\Delta E\F$ of PBE functional is almost zero.
On the other hand, DC-PBE[HF] and HF-PBE lie slightly higher than CCSD(T) and differ from each other.
HF-PBE, DC-PBE[HF], and DC-PBE[CC] become closer to CCSD(T) as the H-Cl distance increases.
So, once again, HF-PBE greatly improves over SC-PBE, but more accurate densities with inversions yield slightly better results. 

\begin{figure} [h!]
\centering
\includegraphics[width=1.0\columnwidth]{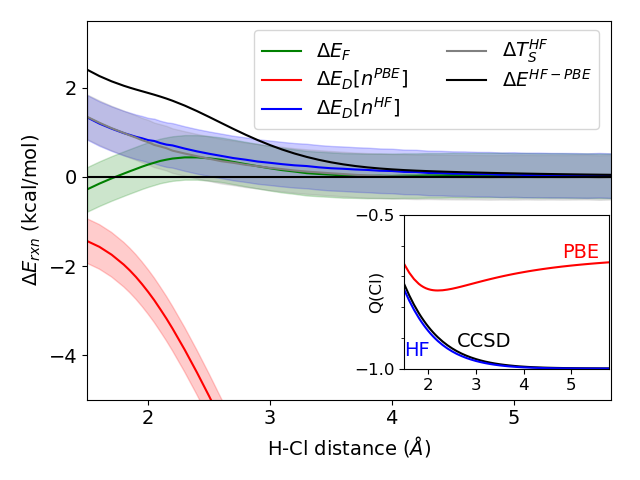}
\caption{
Error components of PBE and $\Delta T\HF\s$ of OH$\cdot$Cl$^-$ dissociation curve.
To represent the uncertainty of the inversion, curves are drawn as $\pm$0.5~kcal/mol bands. 
$\Delta E\D[n\HF]$ and $\Delta T\HF\s$ are almost overlapped with each other.
Inset shows the IAO population of the Cl atom.
The error of HF-PBE is shown in a black solid line with no band because there is no inaccuracy due to inversion.
}
\label{fgr:HOClerr}
\end{figure}

It is noticeable that the partial charges of the Cl fragment for any geometries are similar to the charge at the dissociation limit, as in the inset of Figure~\ref{fgr:HOClerr}.
The behavior of $\Delta E\D[n\PBE]$ and $\Delta E\D[n\HF]$ are similar to that of NaCl.
However, $\Delta T\HF\s$ is not negligibly small for short H-Cl distances; it almost overlaps with $\Delta E\D[n\HF]$ in Figure~\ref{fgr:HOClerr}.
$\Delta T\s\HF$ needs not be zero for any reaction.
The success of HF-PBE requires only that it be much smaller than $|\Delta E\D|$ whenever $|\Delta E\D(PBE)| > $ 2~kcal/mol.

Of course, in most DFT calculations, the error of SC-DFT does not originate from large $\Delta E\D$.
The examples shown here have small $\Delta E\F$, small $\Delta E\D[n\HF]$, and large $\Delta E\D[n\PBE]$, and so are greatly improved by the use of HF-DFT.

\section{conclusion}

We have shown here the reliability of present KS inversion methods for the calculation of the density-driven and functional errors of common KS-DFT.
Some known issues prohibit exact KS inversions in localized basis sets.
From KS inversion methods, ZMP and WY, we show that these issues have a minor effect on reaction energies (sub-kcal/mol), when the inversion is performed with proper conditions; such as an approximate guiding potential, $\lambda$ in ZMP, or potential basis set in WY.
Our recommendations are; \\
1) In the case of ZMP, a KS-DFT guiding potential works better than FA, even for the inversion of a non-KS density.
On the other hand, results are not sensitive to the guiding potential in WY. \\
2) Large $\lambda$ for ZMP: Practically 512 suffices, larger $\lambda$ may lead to convergence issues. \\
3) For the potential basis set in WY, using a basis with the same level of $\zeta$ as the atomic orbital basis in addition to tight-core functions.\\
Under these conditions, one can accurately estimate the density-driven and functional error of common KS-DFT calculations and also estimate the small errors introduced by the HF-DFT procedure. 
We expect that calculation of these errors will help the development of new XC functionals that reduce both $\Delta E\D$ and $\Delta E\F$.

\section*{Computational Details}

Coupled cluster singles, doubles (CCSD) is used as a reference density, while perturbative triples (CCSD(T)) is used as a reference of energy. 
All HF, DFT, and CC calculations were performed using PySCF program package.\cite{PySCF18}
Since we used PBE\cite{PBE96} as the default energy functional, FA, SVWN,\cite{D30, VWN80} and BLYP\cite{B88,LYP88} are tested as guiding potentials for inversions.
The unrestricted scheme is used for open-shell systems.
No frozen-core approximations were made for CC calculations.
The aug-cc-pVTZ atomic orbital basis set is used for both NaCl molecule, and HO$\cdot$Cl$^-$ complex.\cite{D89, D92, D93}
We set gradient converge threshold (conv\_tol\_grad attribute in SCF base class in PySCF) for HF to 1e-7, to generate accurate reference HF determinant for CC calculations.
For the convergence of PBE on the stretched molecules, we set level\_shift=0.2 and conv\_check=False.
All ZMP and WY calculations were conducted with our codes.
For ZMP, we used direct inversion of iterative subspace algorithm\cite{DIIS80} to accelerate convergence.
We solved ZMP equations self-consistently for a $\lambda$ and used the output density matrix as an initial guess of the next ZMP equation with a larger $\lambda$. 
We say ZMP fails to converge at $\lambda'$ when it fails to converge when the initial guess density matrix is from $\lambda'-1$.
For WY, we used the Broyden–Fletcher–Goldfarb–Shanno algorithm\cite{NW06} implemented in SciPy\cite{SciPy} for the optimization of the KS potential. 


\section*{Associated Content}
Tables for all inversion results corresponding to all tables and figures are provided.
Also, we present inversion results of CCSD and CCSD(T) densities for NaCl, and ZMP results for HO$\cdot$Cl$^-$ complex.
Please see supporting information.


\vspace{5mm}
\begin{acknowledgement}
This work at Yonsei University was supported
by the grant from the Korean Research Foundation
(2020R1A2C2007468). KB acknowledges NSF for Grant CHE 1856165.\\
\end{acknowledgement}


\end{document}